\documentclass[utf8]{frontiersSCNS} 

\usepackage{url,hyperref,lineno,microtype,subcaption}
\usepackage[onehalfspacing]{setspace}
\usepackage{color}
\usepackage{ulem}

\newcommand{\editrm}[1]{\textcolor{red}{\sout{#1}}}
\newcommand{\editadd}[1]{\textcolor{blue}{#1}}
\renewcommand{\editrm}[1]{}
\renewcommand{\editadd}[1]{#1}

\newcommand{\editrmtwo}[1]{\textcolor{red}{\sout{#1}}}
\newcommand{\editaddtwo}[1]{\textcolor{blue}{#1}}

\renewcommand{\editrmtwo}[1]{}
\renewcommand{\editaddtwo}[1]{#1}

\def\keyFont{\fontsize{8}{11}\helveticabold }
\def\firstAuthorLast{Azari {et~al.}} 
\def\Authors{A. R. Azari\,$^{1,2*}$, J. W. Lockhart\,$^{3}$ M. W. Liemohn\,$^{1}$ and X. Jia\,$^{1}$}
\def\Address{$^{1}$ Climate and Space Sciences and Engineering Department, University of Michigan, Ann Arbor, MI, United States\\
$^{2}$ Now at the Space Sciences Laboratory, University of California, Berkeley, CA, United States\\
$^{3}$Sociology Department, University of Michigan, Ann Arbor, MI, United States}
\def\corrAuthor{A. R. Azari} \def\corrEmail{azari at berkeley dot edu}

\begin{document}
\onecolumn
\firstpage{1}

\title[Machine Learning: Planetary Space Physics]{Incorporating Physical Knowledge into Machine Learning for Planetary Space Physics} 

\author[\firstAuthorLast ]{\Authors} 
\address{} 
\correspondance{} 

\extraAuth{}

\maketitle

\begin{abstract}

\section{}
Recent improvements in data collection volume from planetary and space physics missions have allowed the application of novel data science techniques. The Cassini mission for example collected over 600 gigabytes of scientific data from 2004 to 2017. This represents a surge of data on the Saturn system. In comparison, the previous mission to Saturn, Voyager over 20 years earlier, had onboard a $\sim$70 kilobyte 8-track storage ability. \editrm{Taking advantage of these data differs from other applications of machine learning due to the unique nature of planetary missions. A primary purpose of applying machine learning to these data is to understand the complexity of planetary systems.} \editadd{Machine learning can help scientists work with data on this larger scale. Unlike many applications of machine learning, a primary use in planetary space physics applications is to infer behavior about the system itself.} This \editadd{raises} \editrm{can be considered through} three concerns: first, the performance of the machine learning model, second, the need for \editrm{explainable and} interpretable applications to answer scientific questions, and third, how characteristics of spacecraft data change these applications. In comparison to these concerns, uses of ‘black box’ \editaddtwo{or un-interpretable} machine learning methods tend toward evaluations of performance only \editadd{either ignoring the underlying physical process or, less often, providing misleading explanations for it}. \editrm{and do not contain evaluations of the inner workings of the application.} \editrm{This work} \editadd{The present work} uses Cassini data as a case study as these data are similar to space physics and planetary missions at Earth and other solar system objects. We build off a previous effort applying a semi-supervised physics-based classification of plasma instabilities in Saturn's magnetic environment, or magnetosphere. We then use this previous effort in comparison to other machine learning classifiers with varying data size access, and physical information access. We show that incorporating knowledge of these orbiting spacecraft data characteristics improves the performance and \editrm{explainability} \editadd{interpretability} of machine learning methods, which is essential for deriving scientific meaning. Building on these findings, we present a framework on incorporating physics knowledge into machine learning problems \editadd{targeting semi-supervised classification for space physics data in planetary environments}. \editrm{We specifically focus on automated detection with space physics data in planetary environments.} These findings present a path forward for incorporating physical knowledge into space physics and planetary mission data analyses for scientific discovery.

\tiny

 \keyFont{ \section{Keywords:} planetary science, automated event detection, space physics, Saturn, physics-informed machine learning, feature engineering, domain knowledge, \editadd{interpretable machine learning}} 

{\noindent{\color{red}Accepted as of May 2020. The typeset version of this article may be found at the Frontiers Research Topic on Machine Learning in Heliophysics under: \href{https://www.frontiersin.org/articles/10.3389/fspas.2020.00036}{ frontiersin.org/articles/10.3389/fspas.2020.00036}.}}

\end{abstract}

\section*{Contribution to the Field Statement}

With the explosion of machine learning usage across scientific fields, a struggle has emerged to derive physical meaning and interpretation from these results. This interest partially stems from a desire to then use these applications to further understanding of complex physical systems. Spacecraft missions offer never before obtained data volumes for characterizing space environments around planets. Spacecraft mission data, however, are inherently challenging to incorporate into machine learning models due to the unique spatio-temporal nature of sampling for instance. Within this work we detail a case study of an \editadd{interpretable} automated event detection method to study Saturn's space plasma environment. We then discuss considerations of machine learning models in a range from physics-informed to physics-blind but data-rich situations \editadd{and provide a framework for future interpretable methods in planetary space physics}. We demonstrate that the incorporation of \editrm{this} \editadd{physical} information improves the performance and usage for scientific understanding, of machine learning methods. This contributes to the fields of space physics and planetary science a path forward to incorporating the domain knowledge of space environments. 

\section{Introduction}\label{sec:Intro}

Planetary space physics is a young field for large-scale data collection. At Saturn for example, it was only in 2004 that the first Earth launched object orbited this planet (Cassini) and landed on Titan (Huygens). After arriving Cassini collected data about Saturn and its near-space environment for 13 years, resulting in 635 gigabytes (GB) of scientific data \citep{NASAJPL2017}. To put this into perspective, the Voyager I mission which flew by Saturn in 1980 had onboard $\sim$70 kilobytes (kB) of memory total \citep{NASA1980}. The Cassini mission represents the first large-scale data collection of Saturn. This enabled the field of planetary science to apply \editadd{statistics to large-scale data sizes} \editrm{large-scale statistics}, including machine learning, to the most detailed spatio-temporally resolved dataset of the planet and its environment.

This surge of data is not unique to Saturn science. In planetary science broadly, Mars in 2020 has 8 active missions roving along the surface and orbiting \citep{PlanetarySociety2020}. The Mars Reconnaissance Orbiter alone has already collected over 300 terabytes (TB) of data \citep{NASAJetPropulsionLaboratory2017}. It is \editadd{commonly accepted} \editrm{expected} that upcoming missions will face similar drastic advances in the collection of scientific data. Traditionally planetary science has employed core scientific methods such as remote observation and theoretical modeling. With the new availability of sampled environments provided by these missions, methods in machine learning offer significant potential advantages. Applying machine learning in planetary space physics differs from other common applications. Cassini's data \editrm{is} \editadd{are} characteristic of other planetary and space physics missions like the Magnetospheric Multiscale Mission (MMS) at Earth and the Juno mission to Jupiter. The plasma and magnetic field data collected by these missions are from orbiting spacecraft. This conflates spatial and temporal phenomena. This is a shared characteristic with the broader field of geoscience which often represents complex systems undergoing significant spatio-temporal changes with limitations on quality and resolution \citep{Karpatne2019}. 

\editrm{The desire to use these data is in order to better understand, or derive fundamental scientific theories.} \editadd{The main use of these data in planetary science is to advance fundamental scientific theories.} This requires the ability to \editrm{back out} \editadd{infer} meaning from applications of \editrm{large-scale} statistical methods. Unlike similar missions at Earth, machine learning for space physics data at Saturn has limited direct application to the prediction of space weather. \editaddtwo{A central interest in space weather prediction is to give lead-time information for operational purposes. As a result, the prediction accuracy in machine learning applications in space weather prediction is seen as paramount.} \editadd{In comparison, at Saturn,} \editrm{As a result,} machine learning applications require highly interpretable and explainable techniques to investigate scientific questions \citep{Ebert-Uphoff2019}. How to improve machine learning generally from an interpretability standpoint is itself an active research area in domain applications of machine learning \citep[e.g.][]{Molnar2019}. \editadd{Within this work we specifically focus on evaluating and implementing interpretable machine learning. Interpretable machine learning usually relies on domain knowledge and is therefore domain specific, but it can be extended to generally refer to models with functional forms simple enough for humans to understand how they make predictions, such as logical rules or additive factors \citep{Rudin2019}. \editaddtwo{Complexity depends in part on what constitutes common knowledge within a domain. Scientists are trained to interpret different models depending on their field. As a result models will range in perceived interpretability across fields.} While the final models must be relatively simple in order for humans to understand their decision process, the algorithms which produce optimal interpretable models often require solving computationally hard problems. Importantly, despite widespread myths about performance, interpretable models can often be designed to perform as well as \editaddtwo{un-interpretable or} ‘black box’ models \editrmtwo{like deep neural nets} \citep{Rudin2019}.} 

In planetary science it's important to discern the workings of a model in order to understand the implications for the workings of physical systems. \editrm{The two terms of interpretability and explainability have differing definitions within the machine learning community. Within this work we will use these two terms interchangeably to represent gaining scientifically actionable results from implementation of machine learning.} \editadd{Interpretability is not the same as explainability: explainability refers to any attempt to explain how a model makes decisions, typically this is done afterwards and without reference to the model's internal workings. Interpretability, however, refers to whether the inner workings of the model, its actual decision process, can be observed and understood \citep{Rudin2019}. Within this work we are concerned with interpretability in order to gain scientifically actionable results from applied machine learning.} The dual challenges of spatio-temporal data and interpretability are compounded for planetary orbiting spacecraft. Complications for orbiting spacecraft can range from rare opportunities for observation, and engineering constraints on spacecraft data transmission. \editadd{A main interest in this work is to begin to ask:} how can machine learning be used within these constraints to answer fundamental scientific questions? 

\editrm{As a potential solution, attention has turned to physics-informed machine learning, of which a primary target has been physics-informed deep learning and discovery of physical concepts (e.g. Raissi et al., 2019; Iten et al., 2020; Ren et al., 2018). Within the space weather prediction community such integration has shown promise in improving the performance of machine learning models (Swiger et al., 2020). These efforts have focused on how to integrate or reveal physical and domain knowledge with machine learning. A long standing interest area is for the discovery of physical laws from machine learning (e.g. Kokar, 1986). For increasing physical understanding, several fields including biology have argued for an equal value of domain knowledge and machine learning techniques (see discussion within Coveney et al., 2016). These discussions have culminated in several reviews for scientific fields on the integration of machine learning for data rich discovery (e.g. Butler et al., 2018; Bergen et al., 2019). Such merged methods underlie a trade between valuing increasing data and model freedom, or incorporating physical insight and model constraint. In Figure 1, we present a diagram for considering physical theory and machine learning within the context of theoretical constraints. The examples at either end of the continuum represent applications of traditional space physics from global theory driven modeling, to space weather and solar flare prediction. The model-adjusted center presented below takes advantage of data, but limits or constrains the application by merging with domain understanding. We argue that working within a model constrained, environment can address both the dual aspects of data characteristics and desired use cases.}

\editadd{Scientists have approached interpretable machine learning for physics in two ways. First, they have added known physical constraints and relationships into modeling. Within the space weather prediction community, such integration has shown promise in improving the performance of deep learning models over models that do not account for the physics of systems \citep{Swiger2020}. Several fields including biology have argued for an equal value of domain knowledge and machine learning techniques for that reason \citep[see discussion within][]{Coveney2016}. These discussions have culminated in several reviews for scientific fields on the integration of machine learning for data rich discovery \citep{Butler2018, Bergen2019}. Second, scientists have long tried to use machine learning for the discovery of physical laws from machine learning \citep[e.g.][]{Kokar1986}. Recently, this work has turned to deep learning tools \citep[e.g.][]{Raissi2019, Iten2020, Ren2018}. However, as \citet{Rudin2019} points out, explanations for the patterns deep learning tools find are often inaccurate and at worst, totally unrelated to both the model and the world it models. These two approaches lie on a continuum between valuing increasing data and model freedom, or incorporating physical insight and model constraint.} 

\begin{figure}[ht!]
    \centering 
    \includegraphics[width=0.8\textwidth]{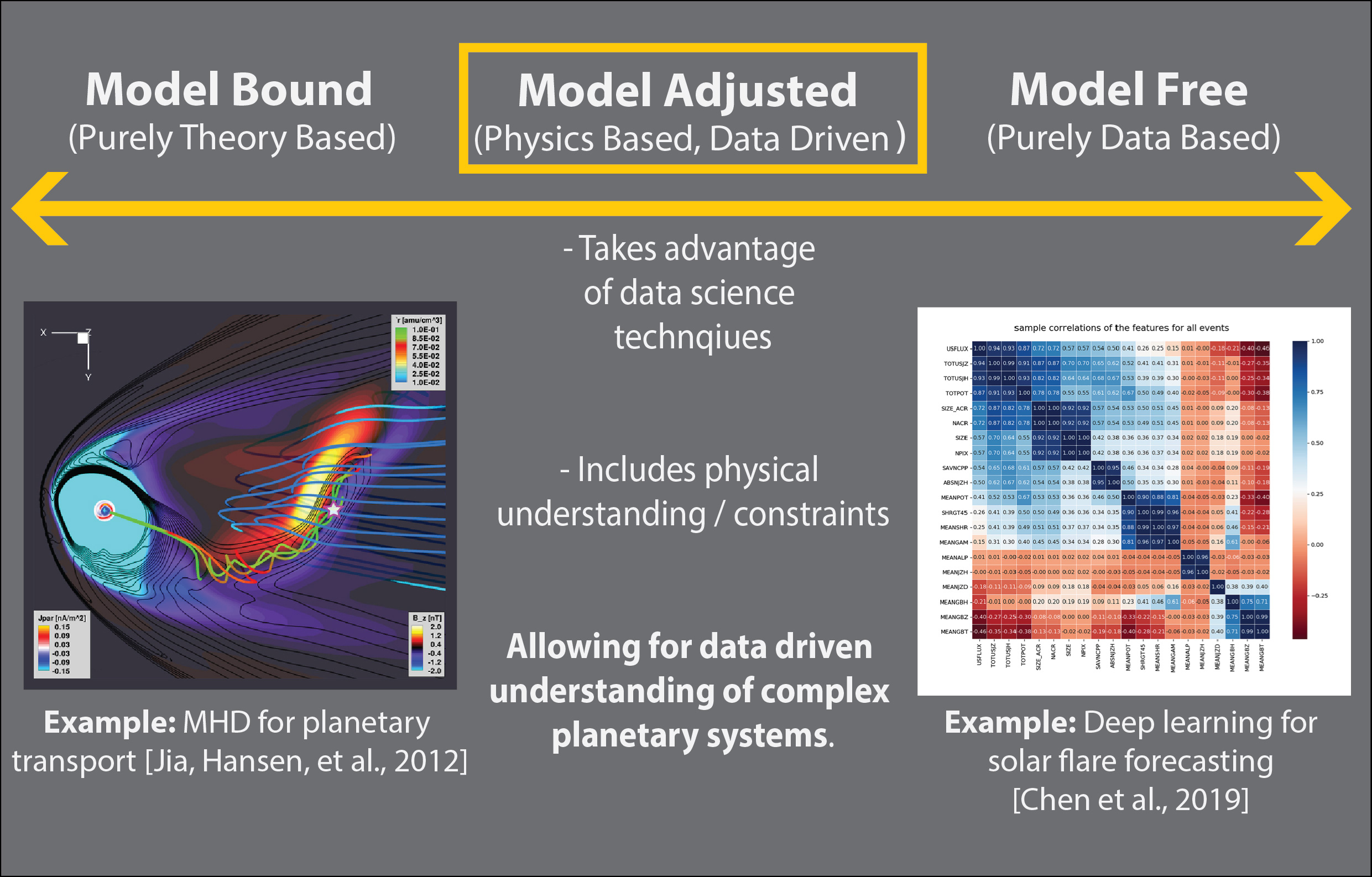}
    \caption[Framework for incorporating physical understanding in machine learning.]{Framework for incorporating physical understanding in machine learning. This figure diagrams a continuum moving from purely theory bound, toward model free. The figure in model bound is from \citet{Jia2012b}, a magnetohydrodynamics simulation of Saturn's magnetosphere. The figure in model free is from \citet{Chen2019}, deep learning feature correlations for solar flare precursor identification. This figure contains subfigures from American Geophysical Union (AGU) journals. AGU does not request permission in use for republication in academic works but we do point readers toward the associated AGU works for citation and figures in \citet{Jia2012b} and \citet{Chen2019}.}
 
    \label{fig:2020MLFig1}
\end{figure}

\editadd{In Figure \ref{fig:2020MLFig1}, we present a diagram for considering physical theory and machine learning within the context of theoretical constraints. The examples at one end of the continuum represent applications of traditional space physics from global theory driven modeling, while those at the other end of the continuum focus on data driven approaches to space weather and solar flare prediction. The model adjusted center presented below takes advantage of data, but limits or constrains the application by merging with domain understanding. Our work is in the middle of the continuum. We leverage domain knowledge about space physics, while also aiming to learn more about the physical system we study. Importantly, we use an interpretable machine learning approach so that we can be more confident in drawing physical insights from the model.} 

\editrm{In the following work} We present comparisons between a range of data sizes and physics incorporation to classify unique plasma transport events around Saturn using the Cassini dataset. As a characteristic data set of space physics and planetary environments, this provides valuable insights toward future implementation of automated detection methods for space physics and machine learning. We focus on three primary guiding axes in this work to address implementations of machine learning. First, we address the performance and accuracy of the application. Second, we consider how to increase \editrm{exoplainability} \editadd{interpretability} of machine learning applications for planetary space physics. Third, we tackle how characteristics of spacecraft data change considerations of machine learning applications. All of these issues are essential to consider in applications of machine learning to planetary and space physics data for scientific interpretation. 

To investigate these questions and provide a path toward application of machine learning to planetary space physics datasets, we compare and contrast physics-based and non-physics based machine learning applications. In Section \ref{sec:Previous}, we discuss the previous development of a physics-based semi-supervised classification from \citet{Azari2018} for the Saturn system within the context of common characteristics of orbiting spacecraft data. We then provide an outline for general physics-informed machine learning for automated detection with space physics datasets in Section \ref{sec:Framework}. Section \ref{sec:Methods} describes the machine learning model set up and datasets that we use to compare and contrast physics-based and non-physics based event detection. Section \ref{sec:Results} details the implementation of logistic regression and random forest classification models as compared to this physics-based algorithm with the context of physics-informed or model adjusted machine learning. Section \ref{sec:Conclusion} then concludes with paths forward in applications of machine learning for scientific insight in planetary space physics.

\section{Background: Saturn's Space Environment and Data}\label{sec:Previous}

\begin{figure}[ht!]
    \centering 
    \includegraphics[width=0.7\textwidth]{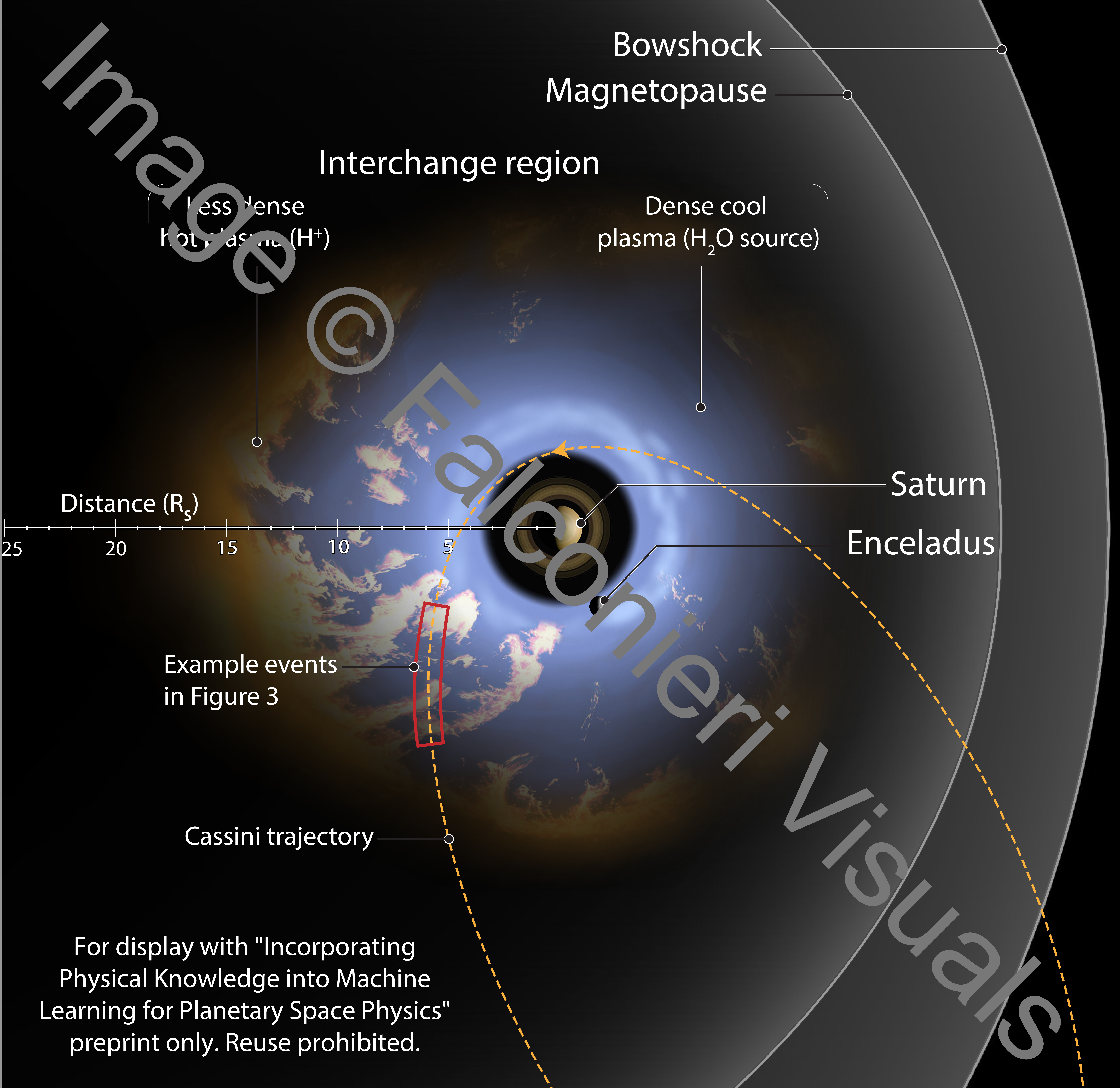}
    \caption[Diagram of interchange injection in the Saturn system.]{Diagram of interchange injection in the Saturn system. The illustrated orbit is an equatorial Cassini orbit from 2005. Injections are denoted by the pale orange material interspersed with the water sourced plasma from Enceladus. Along the example orbit the red box denotes a hypothetical segment of Cassini data discussed in Figure \ref{fig:2020MLFig3}. The purpose of developing an automated event detection is to identify the pale orange material traveling toward the planet. This figure is produced in consultation with, and copyright permissions from Falconieri Visuals.}
    \label{fig:2020MLFig2}
\end{figure}

Saturn's near space environment where the magnetic field exerts influence on particles, or magnetosphere, ranges from the planet's upper atmosphere to far from the planet itself. On the dayside the magnetosphere stretches to an average distance of 25 Saturn radii (R$_{S}$) with a dynamic range between 17 and 29 R$_{S}$ \citep{Arridge2011a} (1 R$_{S}$ = 60,268 km). This distance is dependent on a balance between the internal dynamics of the Saturn system and the Sun's influence from the solar wind. Within this environment a complex system of interaction between a dense disk of neutrals and plasma sourced from a moon of Saturn, Enceladus, interacts with high-energy, less dense plasma from the outer reaches of the magnetosphere (see Figure \ref{fig:2020MLFig2}). 

This interaction, called interchange, is most similar to Rayleigh-Taylor instabilities and results in the injection of high-energy plasma toward the planet. In Figure \ref{fig:2020MLFig2}, a system of interchange is detailed with a characteristic Cassini orbit cutting through the interchanging region. The red box in this figure is presented as an illustrative slice through the type of data obtained to characterize interchange. One of the major questions in magnetospheric studies is how mass, plasma, and magnetic flux moves around planets. At the gas giant planets of Saturn and Jupiter, interchange is thought to be playing a fundamental role in system-wide transport by bringing in energetic material to subsequently form the energetic populations of the inner magnetosphere, and to transport plasma outwards from the moons. Until Cassini, Saturn never had a spacecraft able to develop \editadd{statistics based on large-scale data sizes} \editrm{large-scale statistical data} to study this mass transport system. 

The major scientific question surrounding studying these interchange injections is what role these injections are playing in the magnetosphere for transport, energization, and loss of plasma. To answer this question, it's essential to understand where these events are occurring and the dependency of these events on other factors in the system, such as influence from other plasma transport processes and spatio-temporal location. From Cassini's data, several surveys of interchange had been pursued by manual classification, but these surveys disagreed on both the identification of events and resulting conclusions \citep{Kennelly2013, Lai2016, Chen2008, Chen2010, Muller2010, DeJong2010}. \editrm{This created a need for a standardized survey which was physically justified to allow for subsequent conclusions and comparisons.} \editadd{The main science relevant goal was to create a standardized, and automated, method to identify interchange injections. This list needed to be physically justified to allow for subsequent conclusions and comparisons.}

In Section \ref{sec:PreviousData} we \editadd{provide background on the Cassini dataset} \editrm{discuss the Cassini dataset} and summarize the previous development of a physics-based detection method in Section \ref{sec:PreviousS}. We then provide a generalized framework in the following Section \ref{sec:Framework} for incorporating physical understanding into machine learning with the development of \editadd{this previous physics-based method} \editrm{$S$} as an example. \editadd{Subsequent sections investigate comparisons of this previous physics-based effort to other automated identification methods.} \\

\subsection{Cassini High-Energy Ion Dataset}\label{sec:PreviousData}

\begin{figure}[ht!]
    \centering 
    \includegraphics[width=0.8\textwidth]{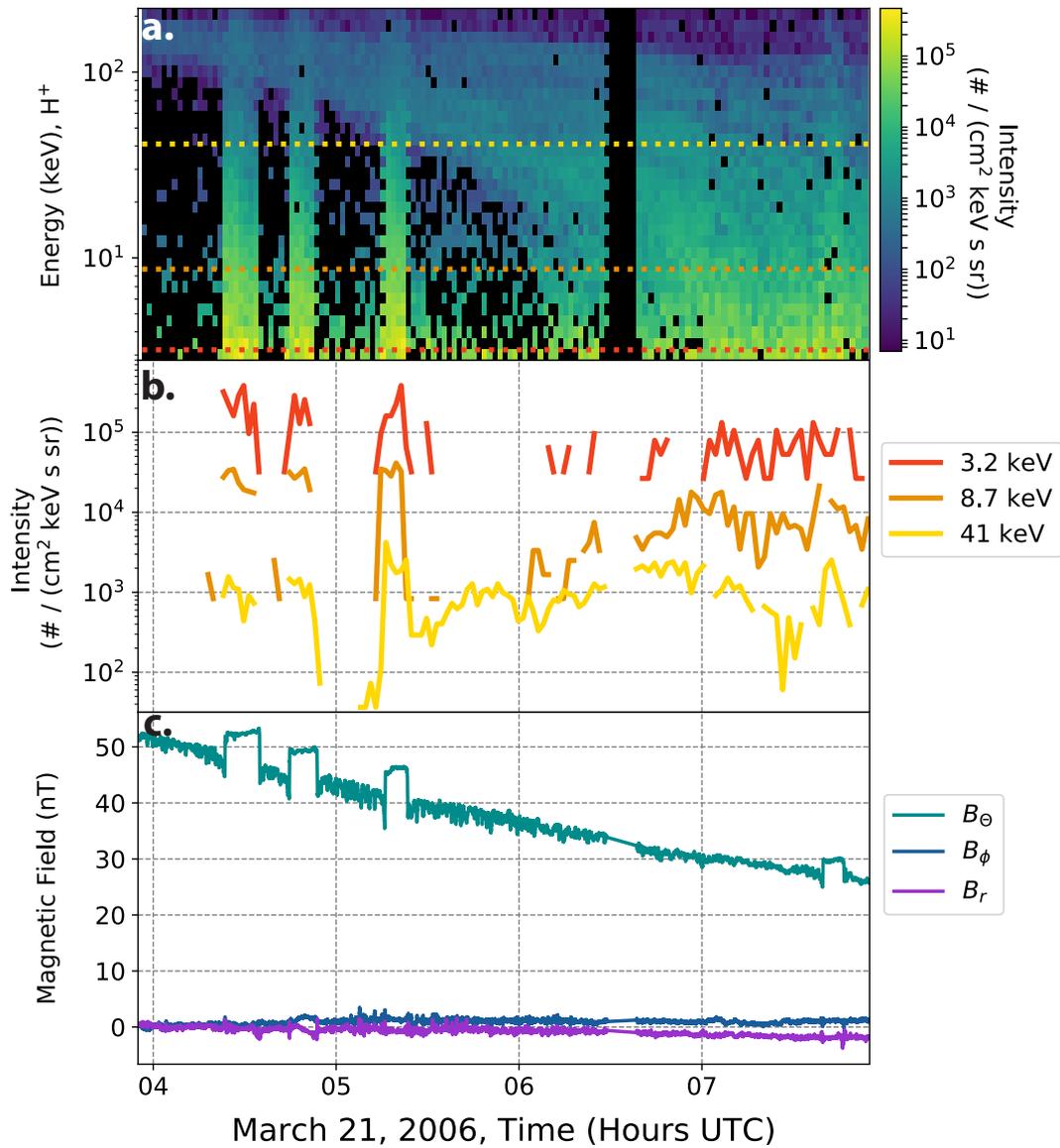}
    \caption[Series of interchange injections characterized by high-energy ions.]{Series of interchange injections characterized by high-energy ions. Panel a details an energy time spectrogram of the intensity from the Cassini CHEMS sensor. \editadd{The color black denotes flux either below the colorbar limit or missing data.}. The three lines are placed at the energy channels for the plot in panel b. Panel b shows the same CHEMS data, but split out into three characteristic energies over the entire CHEMS range. Panel c shows the magnetic field data in KRTP (Kronocentric body-fixed, J2000 spherical coordinates).}
    \label{fig:2020MLFig3}
\end{figure}

Cassini has onboard multiple plasma and wave sensors which are in various ways sensitive to interchange injections. However, none of the previous surveys focused on high-energy ions, which are the primary particle species transported inwards during injections. In Figure \ref{fig:2020MLFig3}\editadd{,} a series of injections are shown in high-energy (3-220 keV) ions (H$^{+}$) and magnetic field datasets. This figure shows three large injections between 0400 and 0600 UTC followed by a smaller injection after 0700 most noticeable in the magnetic field data. \editadd{It is evident from these examples that using different sensors onboard Cassini will result in different identification methods for interchange injections. This was a primary driver in a standardized identification method for these events.} The top two panels detail the Cassini Magnetospheric Imaging Instrument: Charge Energy Mass Spectrometer (CHEMS) dataset while the last contains the Cassini magnetometer magnetic field data \citep{Krimigis2004, Dougherty2004}. 

The CHEMS instrument onboard Cassini collected multiple species of ion data and finds the intensity of incoming particles in the keV range of data. This datastream can be thought of as unique energy channels, each with a spacecraft position and time dependence. In Figure \ref{fig:2020MLFig3}b three unique energy channels are shown from the overall data in the top panel, to illustrate the nature of these high-energy data. This type of spatio-temporal data is often a characteristic of space physics missions \citep[see][for a review of MMS' data products]{Baker2016}. \\

\subsection{Development of Physics-Based Detection Method}\label{sec:PreviousS}
{\editrm{Incorporation of Space Physics into Automated Detection}}

When applying automated or machine learning methods, such data discussed above provides unique challenges and characteristics including: rare events (class imbalances), spatio-temporal sampling, heterogeneity in space and time, extreme high-dimensionality, and missing or uncertain data \citep{Karpatne2019}. These challenges are in addition to desired interpretability. It's essential that an interpretable model is used to learn substantive information about this application. One common use of machine learning is to input \editrm{many values of interest into a black box model. However, as there are many inputs, there are also potentially many relationships within the model.}\editadd{a large number of variables and/or highly granular raw data (e.g. individual sensor readings or image pixel values) into a model, letting the algorithm sort out relationships among them. Such models are inherently ‘black boxes’ because the number and granularity of variables, not to mention complicated recursive relationships among them, makes it difficult or impossible for humans to interpret \citep{Rudin2019}.}  \editrm{A} \editadd{One} solution to this issue is to reduce dimensionality to fewer\editadd{, more meaningful-to-humans} inputs. But at the same time, the model needs to be informative, and the inputs need to be meaningful. Incorporating domain knowledge and then letting the model determine their effectiveness in the system of study is a potential framework to consider. 

For this reason, when developing a detection method to standardize, characterize, and subsequently build off the detected list, a physics-based method was chosen to address these unique challenges. This previous effort is discussed in \citet{Azari2018} and the resultant dataset is located on the University of Michigan's Deep Blue Data hub \citep{Azari2018a}. We build on this effort in the present work to provide a new evaluation of alternative solutions for data-driven methods.

To develop this physics-based method, the common problems in space physics data described in \citet{Karpatne2019} were considered and addressed to develop a single dimension array ($S$). $S$ was then used in a style most similar to a single dimensional logistic regression to find the optimum value for detecting interchange events. This classification was standardized in terms of event severity, as well as physically \editrm{bound} \editadd{bounded} in definition of events. As a result, it was able to be used to build up a physical understanding of the high-energy dynamics around Saturn's magnetosphere including: to estimate scale sizes \citep{Azari2018} and to demonstrate the influence of tail injections as compared to the ionosphere \citep{Azari2019}. \editrm{and derive the role of interchange in the energization and loss within Saturn's mass transport Azari et al., 2020} Following machine learning practices, $S$ was designed through cross validation. It was created to perform best at detecting events in a training \editadd{data}set and then evaluated on a separate test \editadd{data}set. \editadd{These sets contain manually identified events and} \editrm{The test and training set} were developed from 10$\%$ of the dataset (representing 7,375/68,090 time samples). \editadd{Training and test dataset selection and limiting spatial selection is of critical importance in spatio-temporal varying datasets. Our particular selection considerations are discussed in following sections.} The training set was used to optimize the final form of $S$ \editrm{and thresholds of detection}. The test \editadd{data}set was used to compare performance and prevent over fitting. The same test and training \editadd{data}sets are used in the following sections. 

$S$ was developed \editadd{in \cite{Azari2018}} to provide a single-dimension parameter which separated out the multiple dependencies of energy range and space while dealing with common challenges in space physics and planetary datasets. \editadd{$S$ is calculated from $S_{r}$ by removing the radial dependence through normalization.} In mathematical form, $S_{r}$ can be written as: 

\begin{equation}
  S_{r} = \sum_{e = 0}^{14} w^{(Z_{e,r}-C)} \\
  \label{eqn:SValuesML}\\
\end{equation}

\editadd{$S$ can be thought of as a single number which describes the intensification of particle flux over a normalized background. In other words, $S$ can be calculated as: $S = (S_{r} - \bar S_{r}) / \sigma_{S_{r}}$. In which $\bar S_{r}$ is the average radially dependent average and $\sigma_{S_{r}}$ the radially dependent standard deviation. These calculations allow for $S$ to be used across the entire radial and energy range for optimization in units of standard deviation.} \editrm{In the above equation, $S$ is found by combining the normalized values of $S_{r}$ over multiple radial distances between 5 and 12 R$_{S}$}. The variables $w$ and $C$ represent weighting values \editadd{which are optimized for and discussed in the following section}.  The notations of $e$ and $r$ represent energy channel and radial value. $Z_{e, r}$ represents a normalized intensity value observed by CHEMS. \editadd{This is similar to the calculation of $S$ from $S_{r}$}.  

\editrm{$S$ can be thought of as a single number which describes the intensification of particle flux over a normalized background.} Additional details on the development, \editadd{and rationale behind, } $S$ are described in Section \ref{sec:Framework} as a \editadd{specific} example for a general \editrm{strategies toward implementing physical knowledge and this equation is provided as a reference} \editadd{framework for inclusion of physical information into machine learning.} 

The final form of $S$ depends non-linearly on the intensity values of the CHEMS sensor and radial distance. In Figure \ref{fig:2020MLFig4} we show the dependence of the finalized $S$ value over the test \editadd{data}set for the intensity at a single energy value of 8.7 keV and over all radial distances. Within this figure the events in the test \editadd{data}set are denoted with dark pink dots. From  panel d and e it's evident that $S$ disambiguates events from underlying distributions, for example in panel b. By creating $S$ it was possible to create a single summary statistic which separated events from a background population.

The strategies pursued in developing $S$ are most applicable for semi-supervised event detection with space physics data. They can, however, prove a useful guide in starting to incorporate physical knowledge into other applications in heliophysics and space physics. Within the previous effort we used the model optimization process from machine learning to guide a physics incorporated human effort. This was a solution to incorporating the computational methods employed in machine learning optimization to a human-built model. The end result was optimized in a similar fashion as machine learning models but through manual effort to ensure physical\editrm{-} information preservation. Moving from this effort, we now present a framework for expanding the style of integrating human effort and physical-information into other applications for space physics data.   

Below we provide a framework for incorporating physical-understanding into machine learning. In each strategy we discuss common issues in space physics data, using a similar phraseology as \citet{Karpatne2019}. In addition to characteristics in the structure of geoscience data, we also add interpretability as a necessary condition. For space physics and planetary data, the challenges within \citet{Karpatne2019} are often compounded and where appropriate we note potential overlap. After each strategy, we provide a walk-through of the development of $S$ employed in \citet{Azari2018}. \\

\section{Framework for Physics Incorporation into Machine Learning}\label{sec:Framework}

This framework focuses on \editadd{interpretable} semi-supervised event detection with space physics data from orbiters for the end goal of scientific analysis. Depending on the problem posed certain solutions could be undesirable. For a similarly detailed discussion on creating a machine learning workflow applied to problems in space weather, see \citet{Camporeale2019}. The framework presented here can be thought of as a directed application of feature engineering for space physics problems, mostly for requiring interpretability. In general the strategies below provide a context for careful consideration of the nature of domain application which is essential for applications of machine learning models to gather scientific insights.

\begin{figure}[ht!]
    \centering 
    \includegraphics[width=1.0\textwidth]{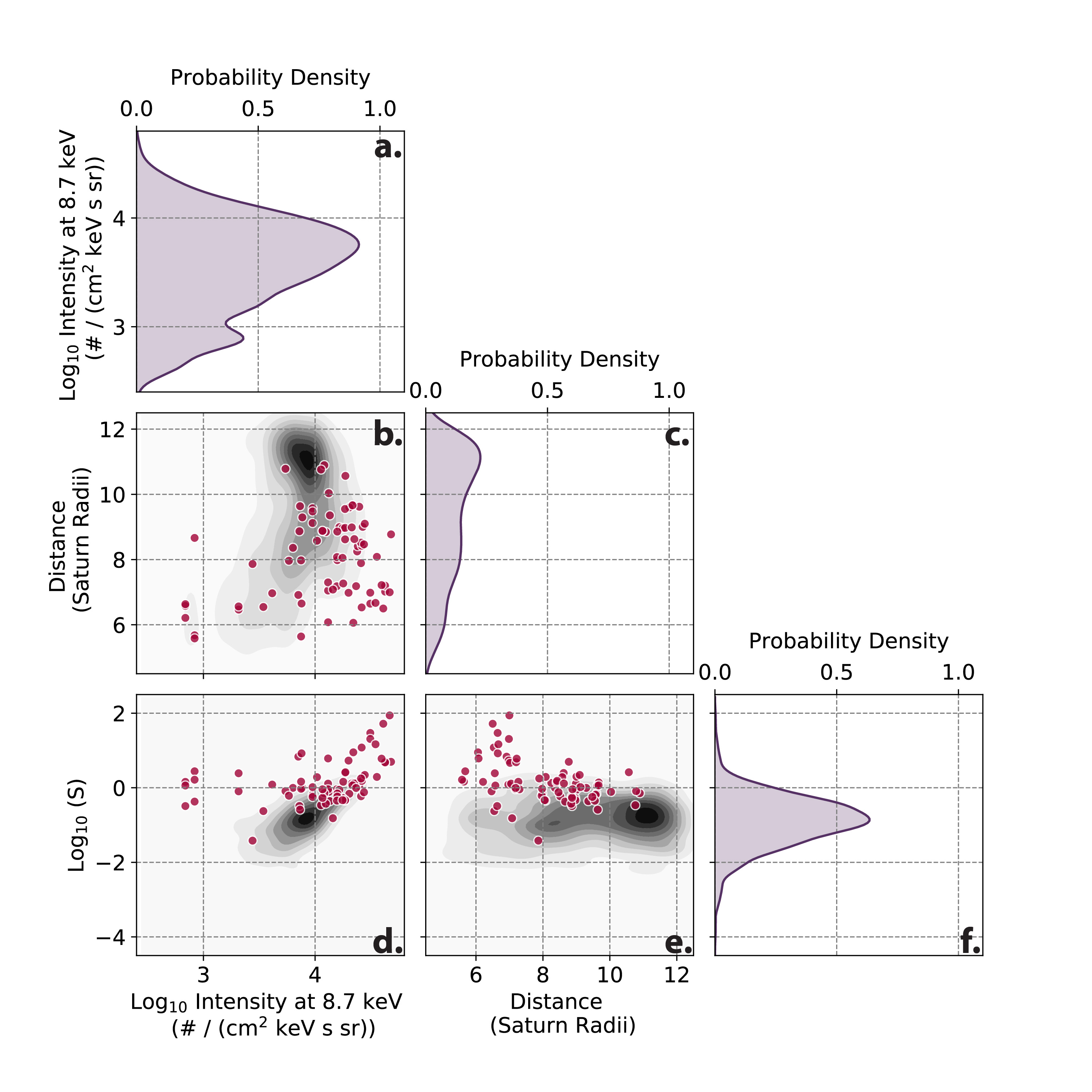}
    \caption[Distributions of$S$parameter developed in Azari et al. 2018.]{Distributions of $S$ parameter developed in \citep{Azari2018}. This figure represents a subset of multiple dependencies of $S$ from a kernel density estimation (kde). The data used in this figure is from the test \editadd{data}set of the data. \editrm{The pink dots within this figure denote the manually identified interchange events within the test dataset.} \editrm{Events in this set are denoted in pink throughout the plot}. Panels a, c, and f represent a single dimension kde of a CHEMS energy channel intensity, spacecraft location in radial distance, and of S. Panels b, d, and e represent two dimensional distributions. This figure was developed using the Seaborn statistics package's kde function \citep{Waskom2020}.}
    \label{fig:2020MLFig4}
\end{figure}

\begin{enumerate}[\bfseries 1.]
    \item{\textbf{Limit to region of interest.} Orbiting missions often range over many environments and limiting focus to regions of interest can assist in automated detection by increasing the likelihood of detection of events. \\
    
    \textbf{Issues}: heterogeneity in space and time, rare events (class imbalance)} \\
    
    \textbf{Example}: The Cassini dataset represents a wide range of sampled environments, the majority of which do not exhibit interchange. In addition, the system itself undergoes seasonal cycles, changing in time, presenting a challenge to any long-ranging spatial or temporal automated detection. The original work targeted a specific radial region between 5 and 12 R$_{S}$ in the equatorial plane. This region is known to be sensitive to interchange from previous studies. Similarly, each season of Saturn was treated to a separate calculation of $S$, allowing for potential temporal changes to the detection of interchange. \\
    
    \item{\textbf{Careful consideration of training and test data\editadd{sets}.} Due to the orbiting nature of spacecraft, ensuring randomness in training and test\editrm{ing} data\editadd{sets} is usually not sufficient to create a representative set of data \editrm{for both sets} across space and time. For event studies, considerations of independence for training and test \editadd{data}set while containing prior and post-event data (at times critical for event identification) are important. This is similar to recent strides in activity recognition studies with spatio-temporal data, in which training set considerations drastically affect the accuracy of activity classification \citep[e.g.][]{Lockhart2014a, Lockhart2014}. \\   
    
    \textbf{Issues}: heterogeneity in space and time, spatio-temporal data, rare events, small sample sizes \\
    
    \textbf{Example}: While the test and training set represent 10$\%$ of the data for the worked example, the 10$\%$ was taken such that it covered the widest range of azimuthal and radial values, while still being continuous in time and containing a range of events.} \\
   
    \item{\textbf{Normalize and/or transform.} Many space environments have a spatio-temporal dependent background. Normalizing separately to spatial or other variables will address these dependencies and can prove advantageous if these are not critical to the problem.\\
    
    \textbf{Issues}: heterogeneity in space and time, spatio-temporal sampling, multi-dimensional data \\
    
    \textbf{Example}: As seen Figure \ref{fig:2020MLFig4}b flux values depend on radial distance and energy value. Similarly, flux exhibits log scaling, where values can range over multiple powers of 10 in the span of minutes to hours as seen in Figure \ref{fig:2020MLFig3}. To handle the wide range of values from the CHEMS sensor, each separate energy channel's intensity was first converted into logarithmic space before then being normalized by subtracting off the mean and dividing by its standard deviation. Effectively, this transforms the range of intensities to a near-normal distribution dependent on radial distance and energy value (see $Z_{e,r}$ in equation \ref{eqn:SValuesML}).} A similar treatment is performed on creating the final $S$ from $S_{r}$. This is important due to the commonality of normalcy assumptions in which models can assume normally distributed data on the same scale across inputs. \\
      
    \item{\textbf{Incorporate physical calculations.} Space physics data can come with hundreds if not thousands of features. While many machine learning techniques are designed for just this kind of data, they do not typically yield results that are amenable to human interpretation and scientific insight into the processes of physical systems. They express a complex array of relationships among raw measurements that do little to help humans build theory or understanding. Summary statistics like summing over multiple variables, or taking integrals, can preserve a large amount of information from the raw data for the algorithm while leaving scientists with smaller sets of relationships between more meaningful variables to interpret. For other fields rich in noisy and incomplete time-series data with a longer history of automated detection methods, summary statistic transformations have been a valuable way of handling this type of data for improved performance \citep[e.g.][]{Lockhart2014a}. \\
    
    \textbf{Issues}: interpretability, multi-dimensional data, missing data \\

    \textbf{Example}: To address missing values \editrm{, not only does building up summary statistics help, but by summing over the energy channels intensity, this creates an integral calculation resembling particle pressure}. \editadd{building up summary statistics, for example through summing over multiple energy channels can help. This creates an particle pressure like calculation (see sum in equation \ref{eqn:SValuesML})}. \editadd{Particle pressure itself is not used to identify events, as the ability to tune the exact parameters was desired in the identification of injections and developing $S$ proved more reliable.} This allows for the lower 14 energy channels to contribute without removing entire timepoints from the calculation where partial data is missing and also increasing interpretability of the end result \editrm{(see sum in equation \ref{eqn:SValuesML})}.} Only the lower 14 channels are used as the higher energy channels also show long duration background from earlier events drifting in the Saturn environment (see Figure \ref{fig:2020MLFig3}).  \\
    
    \item{\textbf{Compare with alternate metrics.} Dependent on your use case, the trade-off costs between false positives and false negatives could be different from the default settings in standard machine learning tools. Investigating alternate metrics of model performance and accuracy are useful toward increasing interpretability. \\
    
    \textbf{Issues}: interpretability, rare events (class imbalance) \\
    
    \textbf{Example}: In the training and test\editrm{ing} \editadd{data}sets only 2.4$\%$ of the data exist in an event state. This proves to be challenging for then finding optimum detection due to the amount of false positives and usage for later analysis. In equation \ref{eqn:SValuesML} scaling factors of $w$ and $C$ are introduced. These scale factors are chosen by optimizing for the best performance of the Heidke Skill Score (HSS) \citep{Heidke1926}. HSS is more commonly used in weather forecasting than in machine learning penalty calculations but has shown potential for handling rare events \citep[see][for a discussion of HSS]{Manzato2005}. In Section \ref{sec:Results} we evaluate how HSS performs as compared to other regularization schemes (final values: $w$ = 10, $C$ = 2).} \\
    
    \item{\textbf{Compare definitions of events, consider grounding in physical calculations}. Much of the purpose of developing an automated detection is to standardize event definitions. Developing a list of events then can become tricky. \\
    
    \textbf{Issues}: lack of ground truth, interpretability, rare events (class imbalance) \\
    
    \textbf{Example}: At this point in the calculation of $S$, there is a single number, in units of standard deviations, for each time point. \editadd{This calculation so far, takes in the flux of the lowest 14 energy channels of CHEMS before normalizing and combining these values to return a single value at each time. This number is higher (in the useful units of standard deviation) for higher flux intensifications and lower for flux drop outs.}
    \editrm{This represents, at its most basic, normalized flux intensifications of the lowest 14 energy channels of CHEMS above the plasma background.} The final question becomes at which $S$ value should an event be considered real or false. \\
    
    Based on the training \editadd{data}set, 0.9 standard deviations above the mean of $S$ is the optimum parameter for peak HSS performance. \editadd{As discussed in Section \ref{sec:PreviousS} 0.9 was determined through optimizing against the training set}. Since $S$ is in terms of standard deviations, additional higher thresholds can be implemented to sub-classify events into more or less severe cases with a physical meaning (ranking). This allowed for the application as a definition task with a physical justification}. \\
    
    \item{\textbf{Investigate a range of machine learning models and datasets.} Incorporating a range of machine learning models, from the most simple to the most complex in addition to varying datasets, can offer insights in the nature of the underlying physical data. \\
     
    \textbf{Issues}: interpretability \\
    
    \textbf{Example}: In developing $S$, alternative feature inclusions were considered. $S$ was settled on for its grounding in physical meaning. A secondary major consideration was its accuracy compared to other machine learning applications. In the following sections we discuss additional models.}
\end{enumerate}

As similarly discussed within \citet{Camporeale2019}, the desire to incorporate physical calculations comes from an interest in using machine learning for knowledge discovery. In the use cases of interest here, both the needs for accuracy and interpretability are essential. These presented strategies are designed to improve the potential performance for semi-supervised classification problems and the interpretability for subsequent physical understanding. Creating the final form of $S$ was a labor intensive process to create and then optimize. Due to $S$'s non-linear dependence on the features shown in Figure \ref{fig:2020MLFig4}, this was a non-trivial task. Similarly expanding $S$ into additional dimensions is challenging. This is where the machine learning infrastructure offers significant advantages as compared to the previous effort. In the following Sections \editadd{\ref{sec:Methods} and \ref{sec:Results}} we discuss alternative solutions to identification of interchange.

\section{Methods: Models and Experimental Setup}\label{sec:Methods}

In the previous physics-based approach, events were defined through intensifications of H$^{+}$ only, allowing for comparisons to other surveys and advancement of the understanding of events. This was a non-intuitive approach as common logic in application of machine learning algorithms suggests that greater data sizes will result in additional accuracy given a well-posed problem. To explore both \editadd{the potential for higher} accuracy as well as interpretability of the application, we compare the performance of two distinct machine learning models with access to varying data set sizes. \editadd{Below we discuss models we use in this comparison effort.} \\ 

\subsection{Models}

Two commonly used machine learning models for supervised classification are logistic regression and random forest classification. Both are considered standard classification models when applying machine learning and performing comparative studies \citep{Couronne2018}. \editadd{While both models can be interpreted by humans, the additive functional form of logistic regression and the broad literature on interpreting it make it highly interpretable. Random Forest models consist of easy to interpret logical rules, but the large numbers and weighted combinations of those rules mean it is less interpretable \citep{Rudin2019}.} The original physics-based algorithm was designed with a logistic regression method in mind, but with significant adjustment. Comparisons to this model are directly informative as a result. Logistic regression categorizes for binary decisions by fitting a logistic form, or a sigmoid. Logistic regression is a simple, but powerful, method toward predicting categorical outcomes from complex datasets. The basis of logistic regression is associated with progress made in the 19$^{th}$ century in studying chemical reactions, before becoming popularized in the 1940s by \citet{Berkson1944} \citep[see][for a review]{Cramer2002}. \editadd{When implemented and optimized using domain knowledge, highly interpretable models, like logistic regression, generally perform as well as less interpretable models and even \editrmtwo{‘black box’} deep learning approaches \citep{Rudin2019}.}

Random forest in comparison classifies by building up collection of decision trees trained on random subsets of the input variables. The predictions of all trees are then combined in an ensemble to develop the final prediction. Similar to logistic regression, the method of random forest has been built over time with the most modern development associated with \citet{Breiman2001}. While logistic regression requires researchers to specify the functional form of relationships among variables, random forests add complexity toward classification decisions, by allowing for arbitrary, unspecified non-linear dependencies between features, also known as model inputs.

The models used within this chapter are from the scikit-learn machine learning package in Python \citep{Pedregosa2011}. Within the logistic regression the L2 \editadd{(least squares)} regularization penalty is applied. Within the random forest a grid search with 5-fold cross-validation is used to find the optimum depth between 2 and 5, while the number of trees is kept at 50. These search parameters are chosen to constrain the random forest within the perspective of the noisy nature of the CHEMS dataset and to prevent over fitting. Alterations to this tuning parameter scheme are not seen to alter the results in the following section. \editrm{Unless
otherwise stated, models are used in conjunction with balanced class weights which adjusts event weighting to be proportional to the frequency of events and non-events.} \editadd{Events are relatively rare in the data (2.4\% of the data in the training and test datasets corresponds with an event), and this can bias the fit of models. As such, unless otherwise noted, we use class weighting to adjust the importance of data from each class (event and non-event) inversely proportional to its frequency so that the classes exert balanced influence during model fitting.} This results in events weighted higher \editadd{more important} than non-events due to their rarity. Performance is shown in Section \ref{sec:Results} against the test \editadd{data}set defined above. \\

\subsection{Dataset Definitions and Sizes}

To explore the performance of logistic regression and random forest, four distinct subsets of the Cassini plasma and magnetic field data are utilized ranging in data complexity and size as follows:

\begin{enumerate}
\item{S{\textbackslash}C (Spacecraft) Location and Magnetic Field

6 features, 68,090 time samples}

\item{S{\textbackslash}C Location, Magnetic Field, and H$^{+}$ flux (3-220 keV)

38 features, 68,090 time samples}

\item{Low Energy H$^{+}$ flux (3-22 keV)

14 features, 68,090 time samples}

\item{Azari et al., 2018 ($S$ Value)

1 feature, 68,090 time samples}

\end{enumerate}

These subsets are chosen to represent additional features, complexity, and physics inclusion. All of these subsets should be sensitive in varying amounts toward identification of interchange injections as evidenced in Figures \ref{fig:2020MLFig3} and \ref{fig:2020MLFig4}. The first two datasets are a comparison of increasing features that should assist in identification of interchange injection. The third dataset includes less features, but is the originator most similar to the derived parameter from \citet{Azari2018}. The final dataset contains the single summary statistic array of the $S$ parameter. In the following result section, these four dataset segments are used to evaluate the two models. 

\section{Results and Discussion}\label{sec:Results}

We are interested in evaluating how the former physics-based $S$ parameter performs with other commonly used subsets of space physics data. \editadd{Our primary goal in this section is to investigate the trade off between the performance of these more traditional models and their interpretability, and therefore usage for scientific analyses.} We complete this through applying supervised classification models and evaluate the ease of \editrm{explainability} \editadd{interpretability} and their relative performance. \\ 

\subsection{Supervised Logistic Regression Classification}

\begin{figure}[ht!]
    \centering 
    \includegraphics[width=0.58\textwidth]{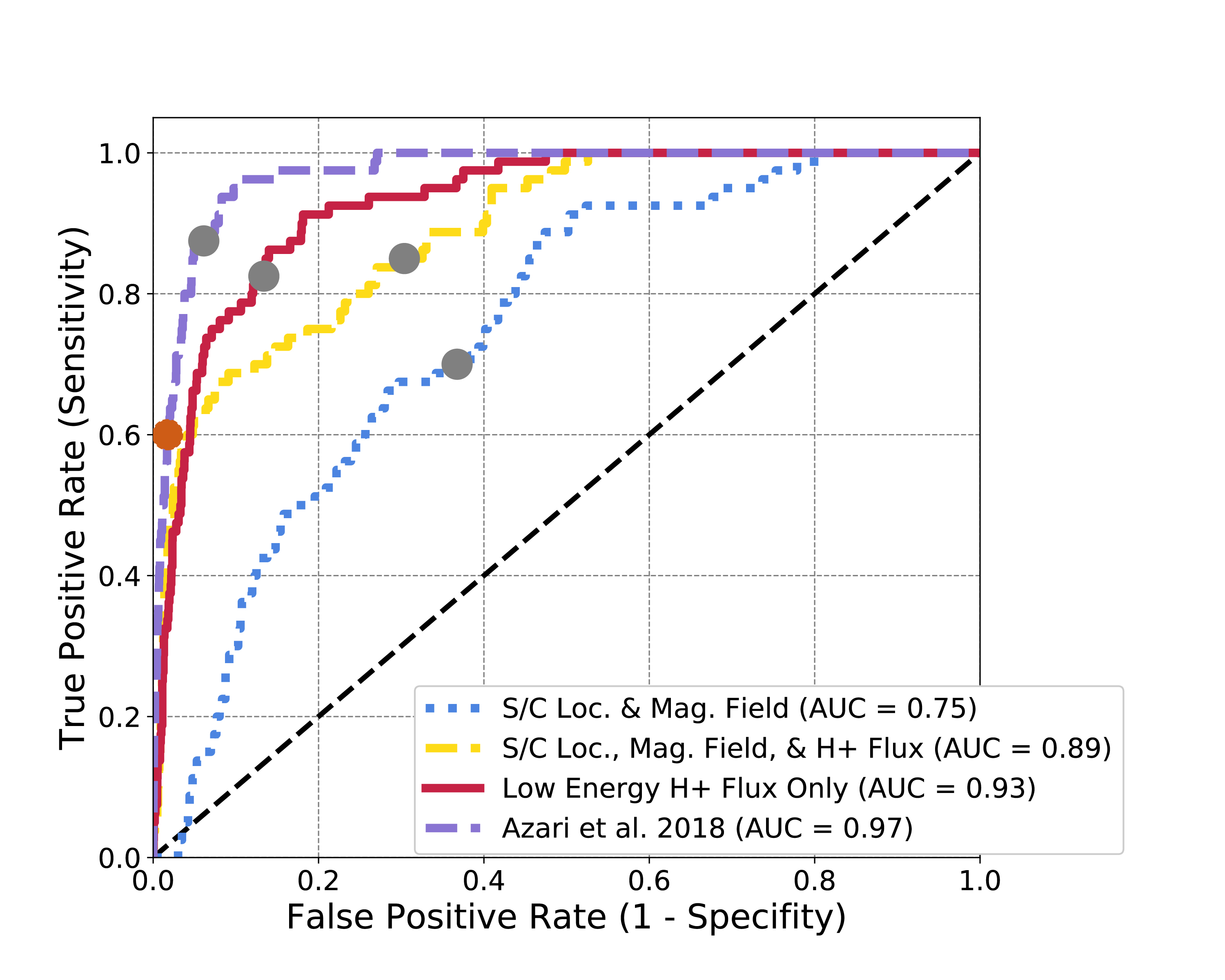}
    \caption[Logistic regression ROC diagram for Cassini data subsets.]{Logistic regression ROC diagram for Cassini data subsets. The grey dots represent the cut-off for L2 regularization for logistic regression. The orange dot represents the peak HSS value, used for optimization in Azari et al., 2018. The distinct curves represent separate ROC curves for each subset of data described in section \ref{sec:Methods}. \editadd{The Azari et al., 2018 subset denotes the usage of $S$.}}
    \label{fig:2020MLFig5}
\end{figure}

In Figure \ref{fig:2020MLFig5} the ROC curve of a logistic regression for all four subsets of Cassini data is presented. \editadd{ROC or receiver operating characteristics, are a common method employed for visualizing the efficacy of classification methods \citep[see][for a generalized review of ROC analysis]{Fawcett2006}. ROC curves in this particular example are created by sweeping over a series of classification thresholds. Ideally a perfect classifier will result in a curve that carves a path nearest to the upper left corner.} Area under the curve, or AUC is presented as a metric to understand the overall performance of each logistic regression evaluation. AUC has the ideal parameters of ranging between 0 and 1, with 0.5 representative of random guessing, 1 representing perfect classification, and 0 as the inverse of truth. AUC can be thought of as an average accuracy of a model and isn't sensitive to class-balance and thresholds. \editadd{ROC curves present the ratios of true positive rate (y-axis) to false positive rate (x-axis). This can be thought of as the trade off for classifiers between events successfully identified (y-axis), and events unsuccessfully identified (x-axis).}

The purple curve represents the logistic regression \editadd{with only the derived physics-based $S$ as an input} \editrm{evaluated with the derived physics-based $S$ described in the above sections}. \editadd{This is rather redundant with optimizing by hand as it's a single variable space. Instead the purple curve is provided as a benchmark against the identical performance and curves found within \citet{Azari2018}.} \editrm{This is provided as a benchmark as it results in the same performance discussed in Azari et al., 2018.} From this figure, this single summary statistic \editadd{($S$)} outperforms all other subsets of Cassini data with an AUC approaching near 1.0 (0.97). \editadd{This is evidence for the current case, that incorporating physical information, even at the expense of greater dataset size improved the performance of certain machine learning applications.}

\editadd{Following this} \editrm{Unexpectedly,} it is not the largest dataset that has the second best performance. Instead, the red curve which contains only the low energy H$^{+}$ intensities shows the best performance of the non physics-adjusted datasets. The magnetic field is a useful parameter for the prediction of interchange as demonstrated in Figure \ref{fig:2020MLFig3} but the form of the logistic regression is unable to use this information successfully. \editadd{This is possibly due to the higher time resolution needed for interchange identification from magnetic field data and any future identification work needs to focus on adjusting the magnetic field inputs and models. The current dataset is processed such that each time point in the CHEMS set is matched with a single magnetic field vector. Normally within interchange analyses, the magnetic field information is of a much higher resolution. It is likely if a study pursued solely magnetic field data of higher time resolution and processed these data to represent pre and post event states dependent on time, the performance of the magnetic field data would be improved.} It's evident from Figure \ref{fig:2020MLFig4} that $S$ exhibits non-linear behavior from the distribution of $S$ on intensity, distance, and energy. Similarly the magnetic field values likely range over a far range due to the background values, that the linear dependency requirements of logistic regression are unable to use this information. Without the flux data especially (the blue curve) logistic regression is unable to predict interchange as compared to the previous physics-based parameter. 

The AUC doesn't capture the entire picture for our interest. While it shows the performance of the algorithm, it contains information for multiple final classifications of events. The grey dots on Figure \ref{fig:2020MLFig5} demonstrates the chosen cut-point for L2 regularization for class weighted events, or the final classification decision for an optimal trade between real events and false events. Within the previous section, the Heidke Skill Score or HSS was discussed as the final threshold separating events from non-events (denoted as the orange dot on Figure \ref{fig:2020MLFig5}). Deciding the threshold of what separates an injection event from a non-event is critical for the implementation of statistical analysis on the results especially in this case, in which non-events outnumber events at a ratio of $\sim$50:1. One solution would be to rank events, in similar style of the previous work of $S$ with categories of events \citep{Azari2018}. \\

\subsection{Rare Event Considerations}

\begin{figure}[ht!]
    \centering 
    \includegraphics[width=0.58\textwidth]{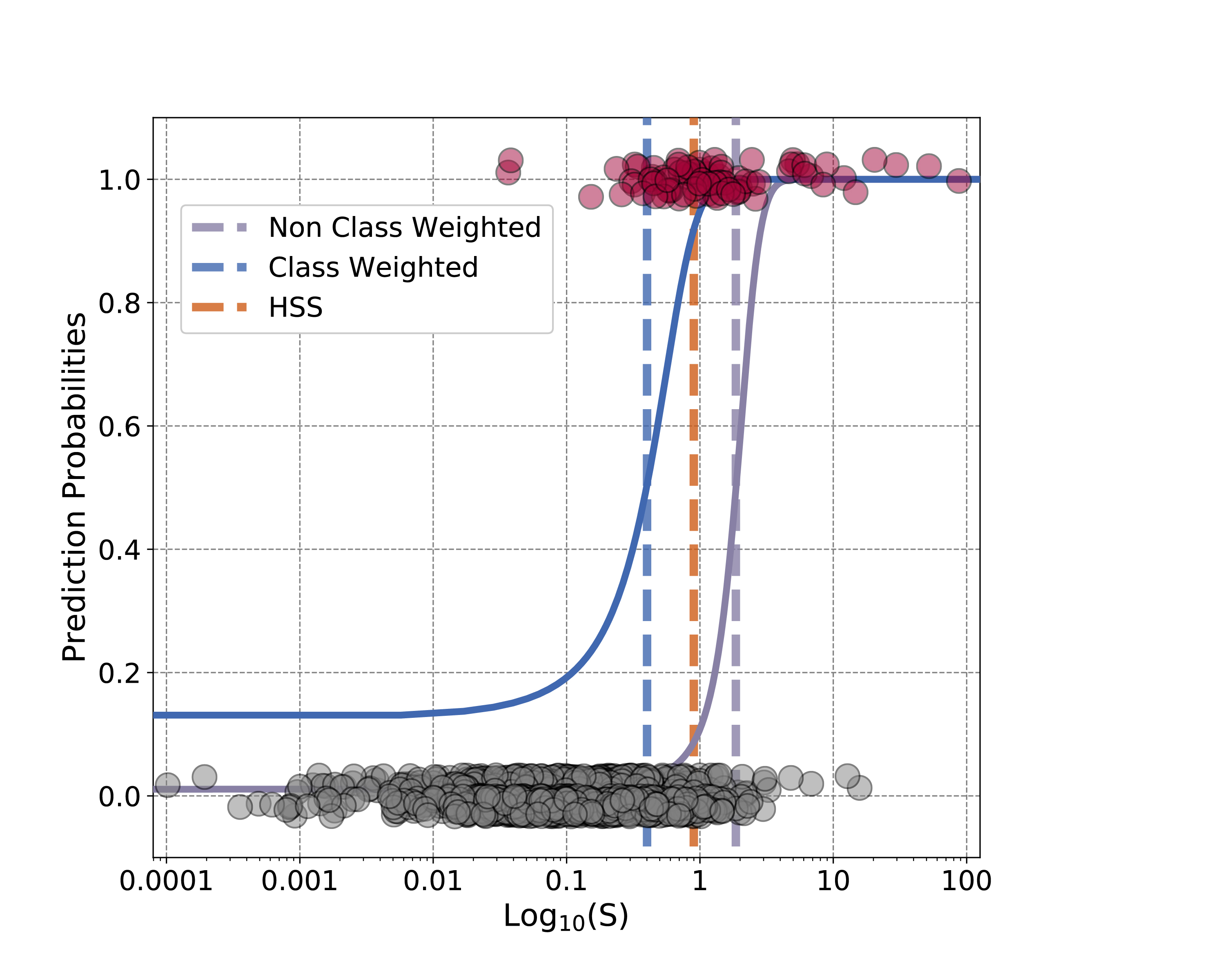}
    \caption[Finalized logistic regression against the test \editadd{data}set \editrm{data}.]{Finalized logistic regression against test \editadd{data}set \editrm{data}. The grey dots represent the test \editadd{data}set values of non-events, and the pink of events. The scatter in the dots around 0 and 1 are for aesthetic reasons and do not represent offset values. This figure contains logistic regressions performed on the physics-based parameter from \citet{Azari2018}. The blue curve represents a class-weighted model and the purple without class weights. \editadd{Similarly the dashed lines for blue and purple represent the finalized cut-off points for the class-weighted and un-weighted models.} The orange dashed line represents the HSS optimization used within \citet{Azari2018}. The x-axis is in logarithmic scale to demonstrate the range of the values, $S$ itself does span both negative and positive values. From being presented in logarithmic space this gives the false illusion that the blue curve does not approach zero.}
    \label{fig:2020MLFig6}
\end{figure}

We now move to evaluating the previous HSS optimization to the logistic regression L2 regulation for both class weighted and non-class weighted models. In Figure \ref{fig:2020MLFig6} the final forms of the weighted and non-weighted logistic regression for the trivial 1 dimensional array case of the $S$ parameter are shown. The thresholds for the final decisions and for HSS are shown as vertical lines \editadd{(the orange dashed line represents HSS)}. Due to the extreme imbalance of non-events to events, implementing class weighting results in large shifts between what is considered an injection event or not. \editadd{We suggest that the class imbalance inherent in this problem is the main rationale between the differences of HSS and other regularizations.} Between the two decision points of the blue and purple vertical lines there are 46 real events, but 202 non-events. This means that if using class-weighting in logistic regression for this problem, 202 non-events would be classified as events. Non-intuitively, for this application where the final events are used to understand the Saturn system, it's advantageous to use a non-class weighted model, as it limits the non-events. However the un-class-weighted model results in removing many real events as well as can be seen in the bulk of the pink events (real events) being misclassified by the purple vertical line. 

The Heidke Skill Score provides an in-between choice of these by providing a higher threshold than the class-weighted, and lower than non-class weighted. The logistic regression for the $S$ parameter shown here is easily intuited since the X-axis represents only one variable. The power of machine learning however is most advantageous in multiple dimensions. HSS has shown to be a more applicable metric for rare events. Other skill scores, such as the True Skill Score have also shown promise in \editaddtwo{machine learning applications to} space physics \citep{Bobra2015}. \editaddtwo{Skill score metrics themselves have a long and rich history in space physics before more recent applications in machine learning with interest originating in space weather prediction \citep[see][for an overarching review of space weather prediction]{Morley2019}.} \editaddtwo{We also direct the reader to discussions of metrics for physical model and machine learning prediction of space weather \citep{Camporeale2019, Liemohn2018}.} How can these traditional metrics for space applications be integrated into the regularization schemes? Future work in machine learning applications should consider shared developments between the physical sciences communities usage of skill scores and regularization of models. \\

\subsection{Supervised Random Forest Classification}

\begin{figure}[ht!]
    \centering 
    \includegraphics[width=0.58\textwidth]{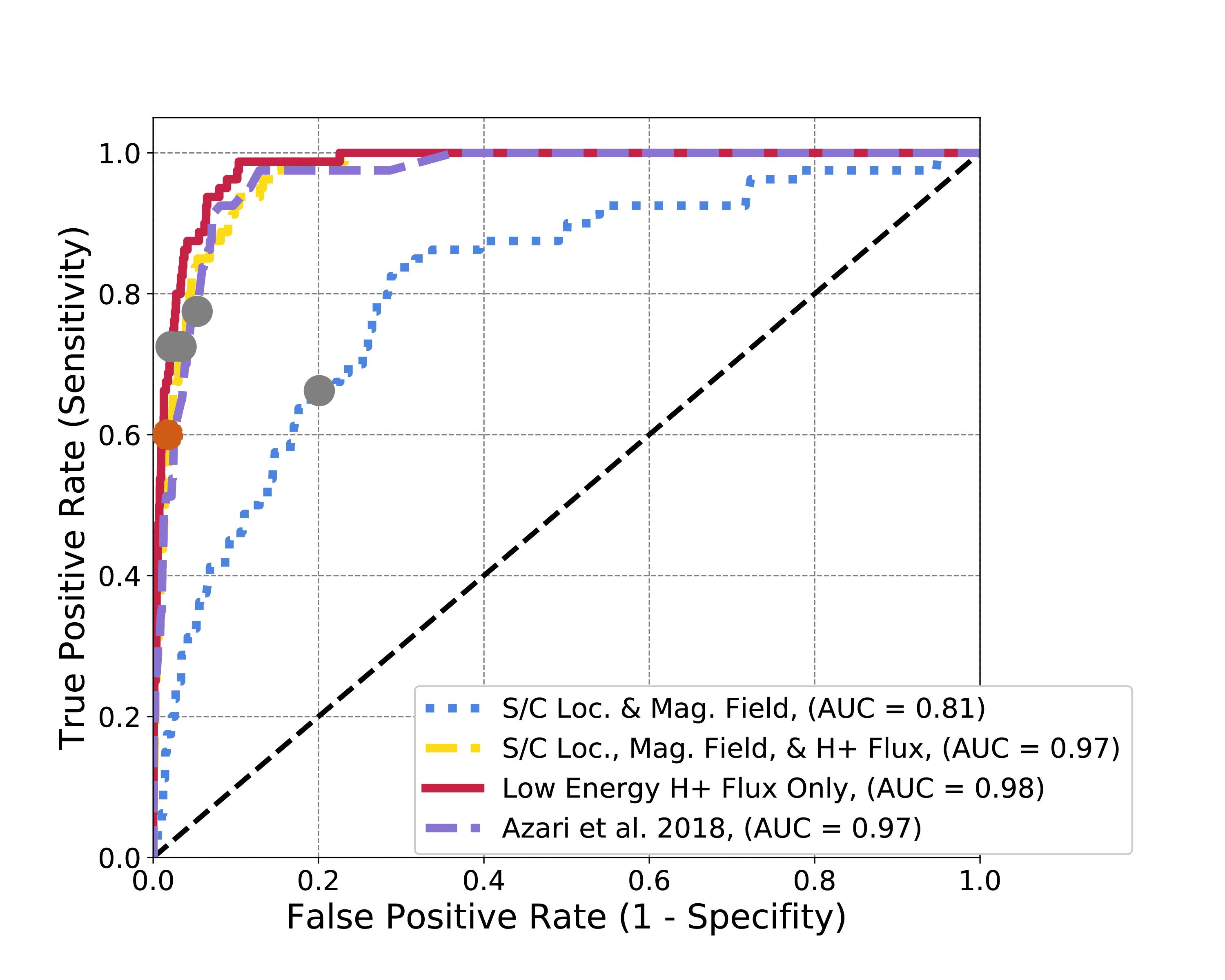}
    \caption[Random forest ROC diagram for Cassini data subsets.]{Random forest ROC diagram for Cassini data subsets. The grey dots represent the final optimization location for random forest classification. The orange dot represents the peak HSS value, used for optimization in Azari et al., 2018. The distinct curves represent separate ROC curves for each subset of data described in section \ref{sec:Methods}. \editadd{The Azari et al., 2018 subset denotes the usage of $S$.}}
    \label{fig:2020MLFig7}
\end{figure}

In Figure \ref{fig:2020MLFig7} the ROC diagram for the same subsets of data but for a random forest model are presented. In this case, unlike the logistic regression, other subsets of data can reproduce the same performance (or AUC) as the derived parameter. All curves, with the exception of the spacecraft location and magnetic field, quickly approach or slightly surpass the AUC of the physics-based parameter at 0.97, with small differences in the performance of the low energy H$^{+}$ flux (0.98) and of the combined spacecraft location, all flux, and magnetic field (0.97). The model form of random forest allows for non-linear behavior in the intensity and magnetic field data to find injection events. Increasing the features then helps in the case of random forest whereas it did not for logistic regression. Similar to the logistic regression, HSS results in a different ratio between true positive rate and false positive rate than the random forest model cut-off point with the grey dots.

Comparing back to logistic regression, even with a relatively complex model such as random forest, the AUC of the best ROC curves are near-identical. Given that $S$ is an array, this is not that surprising. In both cases the physics-derived parameter outperforms or is effectively equivalent to all other data subsets, including those with access to a much richer information set and therefore more complex model. For the application of \editrm{explainability} \editadd{interpretability} for then gathering scientific conclusions, logistic regression is advantageous as it presents a much simpler model. However, random forest, has shown ability to mimic the underlying physics adjustments through selection of datasets. 

Within these results, it's evident that the $S$ parameter performs as well as simplistic machine learning models. Given that $S$ is also grounded in a physics-based definition dependent on solely a variable flux background, this offers advantages to subsequent usage in scientific results. However, many of the adjustments in creating $S$ can be implemented into other space physics data, and integrated into machine learning as evidenced here.  In the description of the development of $S$, several challenges in geoscience data from the framework discussed in \citet{Karpatne2019}, and CHEMS specific solutions were presented. From the above evaluation, it is evident that applications of machine learning are useful to the task of automated event detection from flux data, but with diminishing \editrm{explainability} \editadd{interpretability}. A potential solution to both enhancing the \editrm{explainability} \editadd{interpretability}, similar to the $S$ based parameter, but also incorporating the advantages of machine learning is presented in Figure \ref{sec:Intro}. Rather than consider incorporation of physics-based information as deleterious to the implementation of machine learning, we have found that including this information simplifies the application, enhances the interpretability, and improves the overall performance.

\section{Conclusion and Future Directions}\label{sec:Conclusion}

\editrm{Planetary space physics has reached a data volume capacity at which implementation of machine learning can address scientific questions.} \editadd{Planetary space physics has reached a data volume capacity at which implementation of statistics including machine learning is a natural extension of scientific investigation.} Within this work we addressed how machine learning can be used within the constraints of common characteristics of space physics data to investigate scientific questions.  Care should be taken when applying automated methods to planetary science data due to the unique challenges in spatio-temporal nature. Such challenges have been broadly discussed for geoscience data by \citet{Karpatne2019}, but until now limited attention \editadd{in comparison to other fields} has been given toward \editadd{reviews of} planetary data. 

Within this work we have posed three framing concerns for applications of machine learning to planetary data. First, it's important to consider the performance and accuracy of the application. Second, it's necessary to increase \editrm{explainability} \editadd{interpretability} of machine learning applications for planetary space physics. Third, it's essential to consider how the underlying issue characteristics of spacecraft data changes applications of machine learning. We argue that by including physics-based information into machine learning models, all three concerns of these applications can be addressed. 

For certain machine learning models the performance can be enhanced but importantly in this application, the interpretability improves along with handling of characteristic data challenges. To reach this conclusion we presented a framework for incorporating physical information into machine learning. This framework targeted considerations for increasing interpretability and addressing aspects of spacecraft data into machine learning with space physics data. In particular, it addresses challenges such as the spatio-temporal nature of orbiting spacecraft, and other common geoscience data challenges \citep[see][]{Karpatne2019}. After which we then cross-compared a previous physics-based method developed using the strategies in the framework to less physics-informed but feature rich datasets. 

The physics-based semi-supervised classification method was built on high-energy flux data from the Cassini spacecraft to Saturn \citep[see][]{Azari2018}. In investigating the accuracy of machine learning applications, we demonstrated this physics-based approach outperformed automated event detection for simple logistic regression models. It was found that traditional regularization through L2 penalties both under, and overestimated ideal cutoff points for final event classification (depending on class weighting). Instead, metrics more commonly used in weather prediction, such as the Heidke Skill Score, showed promise in class imbalance problems. This is similar to work demonstrating the applicability of True Skill Score in heliophysics applications \citep{Bobra2015}. Future work should \editaddtwo{consider building on the rich history of prediction metrics in the space physics community for shared development between the physical sciences usage of skill scores and in regularization of models.} \editrmtwo{the shared developments between the physical sciences communities usage of skill scores and regularization of models.} 

While logistic regression is a more interpretable model, random forest proved that with the addition of more \editrm{features} \editadd{and lower level} variables from the Cassini mission, the model could approximate our physics-based \editrm{method} \editadd{logistic model} successfully. In this case physics-informed or model adjusted machine learning, \editrm{as defined within this work,} can \editadd{each} \editrm{be approximated by different models} \editadd{the same performance} but with \editrm{less} \editadd{different levels of interpretability, thus different} ability to \editrm{then make} \editadd{draw} further conclusions about implications of the results. \editadd{The logistic approach provides a coefficient and threshold for a meaningful physical quantity, $S$, effectively the normalized intensification of particle flux. The random forest approach can provide an ‘importance’ score for $S$ or show a large number of conjunction rules involving it, but neither is as useful for human analysts. A forest model using a large number of raw variables instead of a small number of more meaningful ones like $S$ is even harder for humans to make sense of. Deep neural networks, as multi-layered webs of weighted many-to-many relationships, are even less informative for human analysts interested in understanding the workings of the model and physical system. Further, findings that the interpretable model performs as well or better than other approaches demonstrate that, despite the widespread myth to the contrary, there is no inherent tradeoff between performance and interpretability \citep{Rudin2019}.} \editrm{For this application the interpretability is critical.} \editadd{For example, the ability to further split and define identified events based on their flux intensity using $S$ gives the ability to address further scientific questions as to the fundamental mechanisms behind the interchange instability itself.} The simplistic model of logistic regression which results in the same performance as random forest is highly advantageous \editadd{for the current task}. 

The framework and comparison presented here opens up avenues toward consideration of applying machine learning to answer planetary and space physics questions. In the future, cross-disciplinary work would greatly advance the state of these applications. Particularly within the context of interpretability toward scientific conclusions through physics-informed, or model adjusted machine learning. The inclusion of planetary science and space physics domain knowledge in application of data science allows for the pursuit of fundamental questions. We have found that incorporating physics-based information increases the interpretability, and improves the overall performance of machine learning applications for scientific insight. 

\section*{Conflict of Interest Statement}
The authors declare that the research was conducted in the absence of any commercial or financial relationships that could be construed as a potential conflict of interest.

\section*{Author Contributions}
We use the CRediT (Contributor Roles Taxonomy) categories for providing the following contribution description \citep[see][]{Brand2015}. AA led the conceptualization and implemented the research for this manuscript including the investigation, visualization, formal analysis, and original drafting of this work. JL assisted in the conceptualization and discussions of methodology in this work along with editing the manuscript. ML provided funding acquisition, resources, supervision, and assisted in conceptualization along with editing the manuscript. XJ provided funding acquisition, resources, supervision, and assisted in conceptualization.

\section*{Funding}

This material is based on work supported by the National Science Foundation Graduate Research Fellowship Program under Grant No. DGE 1256260 and was partially funded by the Michigan Space Grant Consortium under NNX15AJ20H. JL received funding through an NICHD training grant to the Population Studies Center at the University of Michigan (T32HD007339). ML was funded by NASA grant NNX16AQ04G.

\section*{Acknowledgments}
We would like to thank Monica Bobra, Brian Swiger, Garrett Limon, Kristina Fedorenko, Dr. Nils Smit-Anseeuw, and Dr. Jacob Bortnik for relevant discussions related to this draft. We would also like to thank the conference organizers of the 2019 Machine Learning in Heliophysics conference at which this work was presented, and the American Astronomical Society Thomas Metcalf Travel Award for funding travel to this conference. This work has additionally appeared as a dissertation chapter \citep{Azari2020b}. Figure \ref{fig:2020MLFig2}'s copyright is held by Falconieri Visuals. It is altered here with permission. Figure \ref{fig:2020MLFig1} contains graphics from \citet{Jia2012b} and \citet{Chen2019} which can be found in journals with copyright held by AGU. We would like to thank Dr. Jon Vandegriff for assistance with the CHEMS data used within this work.

\section*{Data Availability Statement}
The events analyzed for this study can be found in the Deep Blue Dataset under doi: 10.7302/Z2WM1BMN \citep{Azari2018a}. The original datasets from the CHEMS \citep{Krimigis2004} and MAG \citep{Dougherty2004} instruments can be found on the NASA Planetary Data System (PDS). Details on the most recent datasets for CHEMS and MAG can be found on the Cassini-Huygens Archive page at the PDS Planetary Plasma Interactions node \href{https://pds-ppi.igpp.ucla.edu/mission/Cassini-Huygens}{(https://pds-ppi.igpp.ucla.edu/mission/Cassini-Huygens)}. Associated data not included in the above repositories can be obtained through contacting the corresponding author.

\bibliographystyle{frontiersinSCNS_ENG_HUMS}

\bibliography{refs}

\section*{Figure captions}



\end{document}